\documentstyle[preprint,aps] {revtex}
\begin{document}

\def\Ni{$Ni^{2+}$}

\draft

\begin{title}
{}~~Understanding Far-Infrared Absorption in the $S=1$
Antiferromagnetic Chain Compound NENP
\end{title}
\begin{center}
Partha P.Mitra  and Bertrand I. Halperin
\end{center}

\begin{center}
Lyman Laboratory of Physics, Harvard University,
Cambridge, MA 02138
\end{center}
\begin{abstract}

Infrared transmission measurements on the $S=1$ antiferromagnetic
chain compound NENP in
applied magnetic fields show a sharp absorption line at the field-shifted
Haldane gap. This violates a wave-vector selection rule of the
Hamiltonian normally used for NENP, as the gap excitations occur
at the Brillouin zone boundary. We argue that the crystal
structure admits terms which can explain the absorption lines.
In addition, in an applied field,
staggered orientations of the g-tensors produce a staggered magnetic field.
This can explain the observation of a finite gap at all applied fields.

\end{abstract}

\narrowtext
The one dimensional $S=1$ antiferromagnet NENP has provided detailed
experimental
 evidence \cite{buyers}-\cite{date} for the current theoretical
 picture \cite{haldane}-\cite{sakai}
 of $S=1$ antiferromagnetic chains, and continues to be the subject of
 a number of investigations.
 However, there remain some puzzling experimental results.
 One of these involves
 infrared transmission measurements \cite{wlu}
in an applied magnetic field in which a sharp absorption line has been
observed at the
field-shifted Haldane gap. This violates the wave-vector selection rule present
in
the simple model Hamiltonian generally employed to describe the system.
In addition,
the
measured gap does not vanish at a critical magnetic field but appears
 to turn around after
reaching a finite minimum value. We propose that the data can be
understood by including in the Hamiltonian terms arising from a staggered
crystal structure which lift the wavevector
selection rule. When a uniform field is applied, the staggered $g-$tensor
effectively produces a staggered magnetic
field, preventing the energy gap from closing.

We begin by comparing the resonance energies from the transmission
experiments \cite{wlu}
with the field-shifted gaps at the zone boundary from
neutron scattering experiments \cite{regnault}
(Figure 1(a)). The transmission experiments are at
a temperature $1.6 K$, much smaller than the gap energy ($14 K$), so
that the absorption is dominated by transitions out of the ground state.
The correspondance of the data demonstrates that light creates
excitations out of the $k=0$ ground state into $k=\pi/a$
excited states at the zone boundary. This violates the
momentum selection rule of the following
Hamiltonian, generally employed for NENP:
\begin{equation}
H=J\sum_i \vec{S}_i\cdot\vec{S}_{i+1} + D\sum_i \bigl(S^z_i\bigr)^2
+E \sum_i \Bigl[\bigl(S^x_i\bigr)^2-\bigl(S^y_i\bigr)^2\Bigr]
-\mu_B\sum_i\vec{S}_i\cdot {\bf g}\vec{H}
\label{eq:orig_ham}
\end{equation}
where $J$ is the \Ni-\Ni exchange, $D$ and $E$ are single ion anisotropies
and ${\bf g}$ is the gyromagnetic tensor. The $z-$axis is chosen along the
chain
axis and $a$ is the spacing between the \Ni ions.
However the correct crystal symmetry of NENP at low temperatures
is given by the space group
$Pn2_1 a$\cite{meyer}. This contains
the symmetry element $2_1$, which is a two-fold screw axis
consisting of a displacement $a$ along the chain, combined with a $180^o$
rotation
about the chain. The screw axis arises from the alternating orientation of the
$NO_2$ groups between the \Ni ions, which is indicated schematically in Fig.2.
The arrangement of $NO_2$ groups on any given chain leads to a uniform
electric dipole
moment along the chain, and an alternating electric dipole perpendicular
to the chain. It also implies that the $D$ and the $g$
tensors of two successive \Ni atoms on a chain have differently
oriented principal axes.
The correct Hamiltonian should therefore be
\begin{equation}
H=J\sum_i \vec{S}_i\cdot\vec{S}_{i+1} + \sum_{i\in (1)}
 \bigl[\vec{S}_i\cdot {\bf D}_1\vec{S}_i-\mu_B\vec{S}_i\cdot {\bf
g}_1\vec{H}\bigr] +
\bigl[\sum_{i\in (2)} \vec{S}_i\cdot {\bf D}_2\vec{S}_i
-\mu_B\vec{S}_i\cdot {\bf g}_2\vec{H}\bigr]
\label{eq:modif_ham}
\end{equation}
where $1,2$ refers to the two
sublattices on the chain, corresponding the two orientations of
 the principle axes of the ${\bf D, g}$ tensors, and we have neglected the
staggered anisotropy of the exchange interaction.
 The presence of the twofold
screw axis implies that ${\bf D_2, g_2}$ are obtained from ${\bf D_1, g_1}$
by a $180^o$ rotation about the chain axis.
As a function of the applied field, (\ref{eq:orig_ham}) and
(\ref{eq:modif_ham})
 have different phase diagrams: for (\ref{eq:orig_ham}),
there exists a critical field strength $H_c$ \cite{affleck} for any
 orientation of the
field,  such that for $H<H_c$ there is no spontaneous staggered
polarisation.
As $H$ crosses $H_c$, there is an Ising transition into a state with staggered
polarisation. For the Hamiltonian (\ref{eq:modif_ham}), the symmetry is
explicitly
broken, and a staggered polarisation exists for any $H\neq 0$.
Moreover, we shall argue below that consistent with the actual symmetry of NENP
and Eq.2, there exists a non-zero electric dipole matrix element leading to
infrared
absorption at the Haldane gap even at $H=0$, whose magnitude could be large
enough
to explain the observations. Magnetic dipole transitions implied by Eq.2 could
also
be important at large $H$.

  The Zeeman term in (2) can be written as
$\sum_i {\vec h}\cdot \vec{S}_i + (-1)^i {\vec h}_{s}\cdot
 \vec{S}_i$ where
${\vec h} = \mu_B{\vec H} ({\bf g}_1 + {\bf g}_2)/2$
 is the effective uniform field and
${\vec h}_{s} = \mu_B{\vec H}( {\bf g}_1-{\bf g}_2)/2$
is the effective staggered magnetic field.
${\vec h}_{s}$ mixes the ground
state with the zone boundary excited states and prevents the energy gap from
closing at a critical field. This staggered field
 has been invoked previously by Chiba {\it et al.}
\cite{chiba}
 to understand the field-dependent splittings of NMR frequencies observed
{\it below} the critical field in NENP.
These authors estimate $\delta g_{yz}/g = 0.01$.

We have employed an approximate model in order to estimate the quantitative
effects of the staggered magnetic field on the triplet energies.
 For simplicity,
we ignore here the staggering of the $D-$tensor.
We model the low lying excitations via a bosonic quantum field theory,
which is modified somewhat from the one proposed by Affleck \cite{affleck}.
In absence of a staggered field, the field-shifted gaps from our
 Lagrangian coincide with results of a fermionic field theory
proposed by Tsvelik \cite{tsvelik}, which appears to give better agreement
with experiment than the unmodified boson theory.
We do not use the fermion theory because there the staggered
polarisation is a non-local operator.
The Lagrangian we use is
($\hbar\equiv 1$):
\begin{eqnarray}
\nonumber
{\cal L} = \int dt\,dx\, &&\Bigl[{1\over 2 v}\Bigl({\partial \vec{\phi}\over
 \partial t}\Bigr)^2
-{v\over 2}\Bigl({\partial \vec{\phi}\over \partial x}\Bigr)^2 -
\sum_i {\Delta_i^2\over 2v}\phi_i^2
+ \sum_{ijl}\epsilon_{ijl} {h_i\over
v}\sqrt{\Delta_j\over\Delta_l}\phi_j\dot{\phi_l}\\
&& +{1\over 2 v}\sum_{ijklm}\epsilon_{ijk}\epsilon_{klm}
\sqrt{\Delta_j\Delta_m\over\Delta_k^2}h_i h_l \phi_j\phi_m -\lambda
(\vec{\phi}^2)^2
-{\bf h}_s\cdot\rho\vec{\phi}\Bigr]~~,
\label{eq:lagrangian}
\end{eqnarray}
where the spin-wave velocity $v$ and  the gaps $\Delta_i$ $(i=1,2,3)$
 at the zone boundary are taken as phenomenological inputs.
We use the values
 $v=110 K$, $\Delta_1=15.7 K$,
 $\Delta_2=13.6 K$, $\Delta_3=29 K$ \cite{buyers},\cite{regnault}.
$\rho$ is a constant relating $\phi$ to the staggered polarisation
$m_s=N^{-1}\sum_{i=1}^N (-1)^i\vec{S}_i$, so that
 $m_s = \rho\phi$. We estimate $\rho$ from the numerically
determined staggered susceptibility \cite{sakai} $\chi_s = 20/J$
($g\mu_B \equiv 1$) for the isotropic model.
 In the isotropic limit of the quadratic Lagrangian,
$\chi_s = \rho^2 v/\Delta^2$. Using $v=2 J$, $\Delta=0.41 J$,
we estimate $\rho = 1.3$. The above Lagrangian
with $h_s=0$, $\Delta_1=\Delta_2$, produces the same field
dependent gaps at the zone boundary as the fermionic
Lagrangian in \cite{tsvelik}.  As an illustration,
for $h_s = 0$, ${\bf h}=h\hat{\bf x}$, the lowest magnon branch has energy
given by
\begin{eqnarray}\nonumber
\omega^2(k)=&&k^2+h^2+{1\over 2}(\Delta_2^2+\Delta_3^2)\\
&&-\sqrt{\Bigl[ \Bigl({\Delta_2^2-\Delta_3^2 \over 2}\Bigr)^2 +
h^2(\Delta_2+\Delta_3)^2+k^2 h^2 \Bigl(\sqrt{\Delta_2\over\Delta_3}+
\sqrt{\Delta_3\over\Delta_2}\Bigr)^2\Bigr]}
\label{example:gaps}
\end{eqnarray}
At $k=0$, this formula is identical to Eq.16c in \cite{tsvelik},
while the formulas differ slightly for finite $k$.
We remark that Date and Kindo \cite{date} have discussed a model where
the magnon energies at the zone boundaries are
those of a single $S=1$ spin
in presence of uniaxial anisotropy. After proper identification of the
zero-field gaps, the field-shifted gaps calculated
from such a single-ion Hamiltonian are identical to
(\ref{example:gaps}) and Eq.16c in \cite{tsvelik}.

To estimate $\lambda$, we need to know the non-linear staggered susceptibility.
In the isotropic limit, the mean field staggered polarisation is
$m_s = \chi_s h_s -
A (\chi_s h_s)^3 + O(h_s^5)$, where $\chi_s = \rho^2 v/\Delta^2$
and $A = 4\lambda v/(\Delta^2\rho^2)$.
{}From perturbation theory, we obtain for large $h_s$,
$m_s=1-(J/h_s)^2 + O((J/h_s)^4)$.
Matching the large and small $h_s$ behaviour using a simple functional form
 $m_s = \chi_s h_s/(1+B \chi_s^2 h_s^2+\chi_s^4 h_s^4)^{1/4}$, we obtain an
estimate
$A=1$,  $\lambda = \Delta^2\rho^2/(4 v)$.
Assuming $\Delta=0.41 J$, $v=2 J$, $J=55 K$ we
have a rough estimate, $\lambda = 3.7 K$.

Finally, we need to estimate the staggered field produced by an
applied uniform field ${\bf h}$. Consider the case where
${\bf h}\parallel \hat{z}$. Taking into account the observations of
Chiba {\it et. al.},  we calculate the field dependent gaps for
(i) ${\bf h} = h\hat{\bf z}$, ${\bf h_s} = 0.01 h \hat{\bf y}$.
A transverse field applied along the $a-$axis makes an
angle of $32^o$ with the $x-$axis of an individual chain. We therefore
calculate field dependent gaps for
(ii) ${\bf h} = h(\hat{\bf x}\cos(32^o)+\hat{\bf y}\sin(32^o))$,
${\bf h_s} = 0.01 \sin(32^o) h \hat{\bf z}$.
The mean field values of the
zone boundary gaps are plotted in Figures 1(a),(b), and
Figure 3 shows the corresponding staggered polarisations.

We now consider the interaction Hamiltonian responsible for the absorption.
The momentum selection rule can in principle be lifted by the presence
of localised impurities or by having chains of finite length.
 However, in this case,
 excited states would be generated with all possible momenta, producing a
threshold instead of a sharp line.
 The observed lines are sufficiently
 narrow so that this seems to be an unlikely explanation.
We will restrict our attention to Hamiltonians that preserve the crystal
symmetry.
 We consider two cases,  (a) interaction Hamiltonians
that violate the translational symmetry of (1), and (b) interaction
Hamiltonians that
maintain the translational symmetry but where
the absorption is allowed in presence of a
staggered polarisation when $H\neq 0$.
In both cases, we consider (i) magnetic and (ii)
electric dipole
transition Hamiltonians. We present below results only for the dc field along
the
chain axis.

In case (a)(i), the transition Hamiltonian is
$H_I(t) = -\mu_B \vec{H_{rf}}(t)\cdot\delta{\bf g}\cdot\sum_i(-1)^i\vec{S}^i$,
where $\vec{H_{rf}}$ is the magnetic field of the infrared radiation.
The absorption intensity for the two lower branches
from  this Hamiltonian should be a maximum when
 $\vec{H_{rf}}\parallel \hat{z}$, and
should vanish for $\vec{H_{rf}}\perp \hat{z}$
 For $\vec{H_{rf}}\parallel \hat{z}$, the absorption coefficient
is
\begin{equation}
\alpha(\omega) = {\mu_0 n (\delta g_{yz}\mu_B)^2 \over \hbar c} \omega
                  \int_{-\infty}^{\infty} {1\over N}\sum_{i,j}
                  \langle (-1)^{i-j}S^i_y(t)S^j_y\rangle e^{i\omega t} dt
\label{abs_coeff}
\end{equation}
where $n$ is the density of nickel ions, $\langle\cdot\rangle$ is the ground
state expectation, and  $\delta g_{yz}/g\approx 0.01$ \cite{chiba}.
At zero applied field, assuming axial symmetry,
$ \langle (-1)^{i-j}S^i_y(t)S^j_y\rangle \approx
(1/3)(e^{-i\Delta_+ t/\hbar}+e^{-i\Delta_- t/\hbar})e^{-\Gamma t} f(|i-j|) $,
 where $f(0)=1$, and $f(i)$
decays exponentially for large $i$ with the transverse correlation length
$\xi$.
 At
resonance, this yields for the lowest branch,
$ \alpha \approx  (3\times 10^{-5}m^{-1}) Q
(\xi/a) \cos^2(\theta_M)$, where $Q=\Delta_+/\hbar\Gamma$ and
$\vec{H_{rf}}\cdot\hat{z} = H_{rf} \cos(\theta_M)$.
We estimate the $Q$ value at low fields
from the experimental data \cite{wlu} to be
be crudely $100$, and also $\xi/a \approx 10$ \cite{buyers}.
 This yields $\alpha \approx 3\times 10^{-2} m^{-1}$. Assuming a sample
 thickness of $1~ mm$, this gives an absorption dip
of $0.003\%$ of transmitted light, small compared to the
observed absorption of $\approx 0.5\% $.

In case (b)(i), the  transition Hamiltonian is
$H_I(t) = -\mu_B \vec{H_{rf}}(t)\cdot{\bf g}\cdot\sum_i\vec{S}^i$.
The orientation dependence of the absorption is the same as in (a)(i).
For $\vec{H_{rf}}\parallel \hat{z}$, the absorption coefficient
can be obtained from Eq.(\ref{abs_coeff}) with the
replacements $\delta g_{yz}\rightarrow g_{zz},~ (-1)^{i-j}\rightarrow 1$.
At low applied fields, noting that
$\sum \vec{S}_i \approx (1/v)\sum \vec{\phi}_i\times \dot{\vec{\phi}}_i$
we find a mean field estimate of the absorption at the lowest gap energy
$ \alpha \approx \alpha_M
Q (\xi/a) (\Delta_+/2 J)^2<\phi_y>^2 \cos^2(\theta_M)$,
where $\alpha_M =  \mu_0 n (g_{zz}\mu_B)^2/\hbar c = 0.3~m^{-1}$
With parameter values as above, this value of $\alpha$ is again too small
 to explain the observed low field intensities.
At high fields, however, the staggered polarisation
becomes quite large. The lowest mode then predominantly consists of angular
fluctuations,
 and we may expect a linearised spin-wave calculation to give qualitatively
correct values of $\alpha$. Such a calculation gives
$ \alpha \approx \alpha_M
Q (2 J\langle\phi_y\rangle/\Delta_+)
\cos^2(\theta_M)$.
The line broadens at high fields, and taking parameter values to be $Q\approx
10$,
$\langle\phi_y\rangle\approx 0.3$, $\Delta_+\approx 5 K$, $\theta_M = 0$, we
obtain
$ \alpha \sim 200 m^{-1}$, implying an absorption
coefficient of
$20\%$ for a sample $1~mm$ in thickness. This is consistent with
the experimental absorption at high fields.

We now consider electric dipole transitions. The \Ni positions
lack a center of inversion symmetry \cite{meyer}, and
 electric dipole transitions are therefore
allowed within the spin manifold. An appropriate Hamiltonian for case (a)(ii)
is a time-dependent Dzyaloshinskii-Moriya (DM) Hamiltonian\cite{moriya}
\begin{equation}
H^{elec}_I = p\vec{E}_{rf}(t)\times \hat{n}_{perp}\cdot\sum_i (-1)^i
         \vec{S}_i \times \vec{S}_{i+1} ~~,
\label{eq:dm}
\end{equation}
where $\vec{E}_{rf}$ is the oscillating electric field from the light wave and
$\hat{n}_{perp}$ is a unit vector along the
staggered component of the
internal electric field $\vec{E_{in}}$.
The static DM exchange
is of the order of \cite{moriya} $J_A\approx J\delta g/g \sim 1 K$. We
write $p\equiv \eta J_A/E_{in}$,
where $\eta$ is a dimensionless constant, and we
 estimate $E_{in}$
 as the field produced by one electronic charge at a distance of $3 A$.
This implies $p = \eta 10^{-33} C-m$.
The absorption in the lowest branches from the above Hamiltonian
vanishes when $\vec{E_{rf}} \perp \hat{z}$
 and is a maximum when $\vec{E_{rf}}\parallel \hat{z}$.
 For the latter case,
\begin{equation}
\alpha(\omega) = {n p^2\over \epsilon_0 \hbar c} \omega
                  \int_{-\infty}^{\infty} {1\over N}\sum_{i,j}
                  \langle A_i(t)A_j \rangle e^{i\omega t} dt
\end{equation}
where $A_i=\hat{z}\times \hat{n}_{perp}\cdot(-1)^i\vec{S}_i \times
\vec{S}_{i+1}$.
{}From the field theoretic description it follows that
$\sum (-1)^i\vec{S}_i \times \vec{S}_{i+1}\approx (1/v)
\sum \partial \vec{\phi}_i/\partial t$. Noting this relation,
we estimate $\alpha$ for the gap energy to be
$ \alpha \approx \alpha_E
Q (\xi/a) (\Delta_+/2 J)^2\cos^2(\theta)$, where
 $\vec{E_{rf}}\cdot\hat{z}=E_{rf} \cos(\theta_E)$
and $\alpha_E = {n p^2/\epsilon_0 \hbar c} = 4 \eta^2 \times 10^{-3}  m^{-1}$.
Taking typical values at low fields to be $Q\approx 100$, $\Delta_+\approx 10
K$
 and $\xi/a\sim 10$, we obtain $\alpha\sim 0.04 \eta^2 m^{-1}$.
In order to obtain an absorption comparable to the experimental value of
$0.05\%$ for a $1~mm$ thick sample, we must take $\eta = 10$, which is large,
 but does not
seem impossible. At high fields, a linearised spin wave theory
yields for the lowest mode $\alpha \approx
{\alpha_E Q }(2 J/\Delta_+)\langle\phi_y\rangle^2\cos^2(\theta_E)$.
For $Q\approx 10$, $\eta = 10$,
$\langle\phi_y\rangle\approx 0.3$, $\Delta_+\approx 5 K$, $\theta_E = 0$, we
obtain
an absorption coefficient of
$10\%$ for a sample $1mm$ in thickness, comparable to
the experimental absorption at high fields.

The Hamiltonian for case (b)(ii) includes
a time dependent symmetric exchange and
a time dependent on-site anisotropy. The absorption from these terms
have the same selection rules at high fields as the case (a)(ii).

To summarise our observations on the absorption intensities: while these do not
strictly vanish
for any orientation of the polarisation plane of light, the magnetic dipole or
electric dipole absorption intensities for a particular line
can individually vanish for special
orientations of the polarisation plane. The magnetic dipole absorption at low
fields
is estimated to be too weak to explain the observations, while the spin wave
estimate at high field for the lowest branch produces numbers closer to the
observation. We cannot estimate the electric dipole absorption with accuracy,
but
our crude estimate indicates that it can produce plausible absorption strengths
at all the applied fields.
A priori, there is no reason why
the two absorption mechanisms should have the same strength,
 so that one might expect significant angular dependence
in the intensities. Careful studies of the angular
dependence of the intensities should clarify the
absorption mechanisms. The strong increase in the absorption intensities
at high fields is presumably related to
the appearence of a large staggered moment.

In conclusion, the experimentally observed infrared absorption in NENP at low
temperatures\cite{wlu}
can be understood after taking into account the crystalline
symmetry of the material. The presence of a staggered, anisotropic
 g-tensor produces a
staggered internal field when a uniform magnetic field is turned on. This
prevents
the gap from closing at a critical field. Mean field calculations of
the gap energies
qualitatively agree with the experimental observations.
Transition hamiltonians can be found consistent with the reported
crystalline symmetry of NENP that break the translational symmetry of the spin
chain, and can explain the observed transitions.

The conclusions of our work are corroborated by recent spin resonance
experiments \cite{brunel}
 where the temperature was varied. The absorptions corresponding to
those described in
\cite{wlu}
rapidly disappear with increasing temperature,
 consistent with the depopulation of the ground state.
As the temperature is raised, absorptions corresponding to flipping the
spins of excited magnons increasingly become visible.
 One remaining mystery about the
absorption, however, is the splitting of each absorption peak
 observed in the experiments of Ref. \cite{wlu}.

We would like to thank D. Huse,
S. Geschwind, W. Lu and I. Silvera for illuminating discussions.
This work was supported in part by NSF grant DMR 91-15491 and by a
fellowship to PPM from the Schlumberger Corporation.

\figure{In (a), the field shifted gaps from neutron scattering (circles)
for ${\bf h}\parallel\hat{z}$ are
compared to the positions of the absorption dips in the transmission
experiment (crosses) .
 (b) corresponds to field applied transverse to the chain, for which only
neutron scattering data (circles) is available.
The solid lines in (a),(b) result from the mean field calculation described in
the text.}
\figure{Schematic diagram of the backbone of an NENP chain,
showing its staggered structure.}
\figure{Mean field staggered polarisations corresponding to the gaps in
Fig.1. The NMR lineshifts from Chiba {\it
et.al.},
are shown as dots. The lineshifts have been scaled by a constant to make one
point lie on the theoretical curve.}

\end{document}